\documentstyle[11pt,aaspp4,flushrt]{article}
\def\etal{\it et al. \rm }
\begin{document}

\title{Gas Rich Dwarfs from the PSS-II I. Catalog 
and Characteristics}

\author{James M. Schombert} \affil{Department of Physics, University of
Oregon}

\author{Rachel A. Pildis} \affil{Harvard-Smithsonian Center for
Astrophysics}

\author{Jo Ann Eder} \affil{Arecibo Observatory\altaffilmark{1}
\altaffiltext{1}{The Arecibo Observatory is part of the National Astronomy
and Ionosphere Center which is operated by Cornell University under
contract with the National Science Foundation.}}

\begin{abstract}
This project is a visual search for field dwarf galaxies using Second
Palomar Sky Survey photographic plates.  A morphologically selected sample
of 310 objects yielded 145 detections of true dwarfs within a redshift
search window of 0 to 10,000 km sec$^{-1}$.  We confirm the low mass,
dwarf nature of the same by comparison of luminosity, isophotal size, HI
mass and HI profile width distribution of other dwarf samples.  The goal
of this project is to use these newly discovered dwarf galaxies to map
large scale structure as a test of biased galaxy formation.  Initial
indicators are that the large scale distribution of dwarf galaxies is
identical to bright, high mass galaxies, in contradiction to theory using
biasing.  The full analysis of the sample will be reported in the final
paper of our series.
\end{abstract}
\section{Introduction}

Dwarf galaxies have received a great deal of observational attention in
the last decade for two reasons.  The first is that their intrinsic
properties provide insight into galaxy formation because dwarf galaxies
represent the other extreme in the global galaxy mass distribution.
Studying the range of galaxy types from dwarfs to giants is a key test of
galaxy formation and evolution models.  The second reason deals with the
capability of dwarf galaxies to serve as tracers of dark matter.
Investigations of dark matter in recent years have questioned the basic
premise that light traces the mass on large scales (see Oemler 1988) and
there have been suggestions that dwarf, or low mass galaxies, may better
sample the distribution of mass in the Universe (Dekel and Rees 1987).
The most extreme views question whether the galaxy formation process
itself may be biased such that galaxies which formed from low amplitude
fluctuations in the early Universe (i.e. dwarf galaxies) are the true
indicators of the large scale distribution of mass (Giovanelli, Haynes and
Chincarini 1986, Davis and Djorgovski 1985).

The hypothesis that bright, or high mass, galaxies do not trace the true
mass distribution of the Universe has come about due to the failure of
various cosmological models to correctly predict the amount of large scale
structure (voids and walls), or the observed peculiar velocities of the
bright galaxies.  Parallel to these efforts was the suggestion from grand
unified theory (GUT) of new types of stable particles with non-zero mass
that interact only weakly with baryons (Turner 1987).  The introduction of
cold dark matter (CDM) resolves the following cosmological problems:  1)
it allows $\Omega_o$=1 without violating the baryon density
($\Omega_b$=0.2) set by primordial nucleosynthesis, 2) CDM allows growth
of present day large scale structure from fluctuations currently measured
by COBE in the cosmic microwave background (CMB), since the CMB only
responds to baryons, 3) CDM can be more smoothly distributed than baryons
such that dynamical estimates for $\Omega_o$ are too low, 4) it resolves
conflicts between cluster models and the galaxy distribution since
galaxies do not trace the mass and 5) weak interactions form the basis to
segregate baryonic and non-baryonic matter to form dark halos (see Oemler
1988 for a review).  The only missing piece is a mechanism to segregate
CDM from baryonic matter.  On small scales, such as galaxy halos,
dissipation is sufficient, but on large scales the mechanism remains
undetermined.  This has led to the speculation that an inefficiency in
galaxy formation produces a bias in the distribution of bright galaxies
such that they only trace the peaks of the mass distribution.  The
theoretical justification for biasing arises from the assumptions that if
galaxies form from high $\sigma$ fluctuations and if those fluctuations
are superimposed on uncorrelated, random, larger-scale fluctuations of
small amplitude, then galaxies will be more clustered than the underlying
matter distribution (Kaiser 1984).

The theoretical community has also found support for a biased galaxy
formation scheme in the many selection effects inherent in galaxy
catalogs.  Selection effects towards high mass galaxies that would
maximize a biased interpretation from observational results such as the
autocorrelation function for galaxies.  For example, there is a clear
surface brightness bias in galaxy catalogs to select against objects with
central surface brightnesses near or below the natural sky brightness.
Theory predicts that low surface brightness (LSB) galaxies are the result
of lower amplitude fluctuations and that redshift surveys, without a
complete sample of LSB objects, are biased (Mo, McGaugh and Bothun 1994).
Although not all LSB galaxies are dwarfs (Bothun \etal 1987) nor are all
dwarfs of a LSB nature (Loose and Thuan 1986), a majority of HI-rich
dwarfs have central surface brightnesses below 23 $B$ mag arcsec$^{-2}$
(Schneider \etal 1990).  Magnitude limited catalogs are known to be sparse
in LSB galaxies (Sandage and Tammann 1981) and angular-limited catalogs,
such as the UGC (Nilson 1973), that have deeper surface brightness levels
will fail to find small dwarf galaxies at useful distances to study large
scale structure since they quickly fall below the angular size limits
for redshifts greater than 1000 km sec$^{-1}$.  Searches for dwarf
galaxies have primarily focused on nearby rich clusters, such as Virgo or
Fornax (Sandage and Binggeli 1984, Caldwell and Bothun 1987), due to the
vast number of objects located in a small region of the sky and with
distances being determined by association with the cluster.  But cluster
catalogs are not useful in studying the large scale distribution of dwarfs.
Current all sky catalogs sample the less dense field population (e.g. the
DDO catalog, van den Bergh 1960) and are rich in the lowest and faintest
type of dwarfs, but are restricted to very local objects.  The purpose of
this project is to overcome these complimentary deficiencies in galaxy
catalogs by attempting to recover field LSB dwarf galaxies through the use
of Second Palomar Sky Survey plates.

In addition to the interest in the properties and content of dwarf
galaxies with respect to global issues of galaxy formation and evolution,
dwarf galaxies are also the leading candidates for the well known faint
blue galaxies excess at moderate redshifts (FBE, see Kron 1980, Tyson
1988, Lilly, Cowie and Gardner 1991).  Current models indicated that a
majority of the FBE can be explained by a simple, evolving dwarf
population (Driver \etal 1996).  Estimates are that up to 30\% of the
galaxy luminosity density is due to dwarfs and the number density is 20
times greater than that of normal or giant galaxies.  This population is
typically low in mean surface brightness and fails to be included in
catalogs based on isophotal magnitude or diameter.  Thus, a secondary goal
of this project is to investigate the optical and HI properties of a
unbiased sample of dwarfs, selected by morphology and unrestricted by
surface brightness. Papers II and III in this series will deal with these
topics.

All distance related values in this paper use values of $H_o = 85$ km
sec$^{-1}$ Mpc$^{-1}$, $\Omega_o$ = 0.2, a Virgo velocity of 977 km
sec$^{-1}$ and a Virgo infall of 300 km sec$^{-1}$.

\section{A Search for Dwarf Galaxies}

The key phase of this project was the identification of objects to test
the biased galaxy formation hypothesis.  For this test, the selection of a
large sample of low mass or dwarf galaxies is required.  There have also
been recent arguments in the literature that low surface brightness
galaxies are better test particles for biasing schemes (see Mo, McGaugh
and Bothun 1994).  In either case, the current catalogs which contain a
suitable number of objects, such as the DDO (van den Bergh 1960) or the
UGC (Nilson 1973), have insufficient depth in redshift space to sample a
significant number of either dwarf or LSB galaxies.

A search for dwarf galaxies has one advantage over other catalog work; the
unique morphology of dwarf galaxies makes them easily recognizable even
at large distances.  Dwarf galaxies come in two general morphological
types, dwarf ellipticals (dE) and dwarf irregulars (dI), basically
reflecting symmetry in their isophotes.  The dwarf ellipticals, although
having the same symmetry as giant ellipticals, differ in their intrinsic
profile shapes having exponential rather than power-law distributions.
This produces a diffuse appearance for dE's on photographic plate material
which easily separates them from background normal ellipticals (see
discussion in Binggeli, Sandage and Tammann 1985).  Dwarf irregulars range
from the extremely high surface brightness blue compact dwarfs (BCD's) to
the very faint Virgo irregulars (see Sandage and Binggeli 1984).  However,
despite the range in mean surface brightness, their chaotic appearance
distinguishes them from every other Hubble type.

As shown in Schombert \etal (1992), a catalog of LSB galaxies is composed
primarily of three classes of objects; 1) dwarfs, 2) LSB disks (quiescent
counterparts of late-type star-forming spirals) and 3) Malin objects
(supergiant LSB disks).  To avoid the latter two categories in our sample,
we concentrated our follow-up observations on objects that appeared to be
Sm, Im, Irr, BCD, dI or dE, with a special emphasis on LSB objects and
objects with small (down to 20 arcsecs) angular size.  We have followed
the prescription for morphological classification as outlined in Sandage
and Binggeli (1984) from the Virgo Cluster catalog.  Although our sample
is a field sample, there is no indication that field irregulars differ in
morphological types from cluster irregulars.

A smaller angular size for this study, as compared to previous LSB
catalogs from PSS-II plates, is key to the investigation of large scale
structure.  For comparison, the UGC, with its one arcmin size limit, would
catalog objects 4 kpc in diameter out to only 1200 km sec$^{-1}$.  Our
visual search produced 310 candidates off of 35 plates (1400 square
degrees).  After eliminating duplication from objects in the overlap
regions between plates, or objects that already exist on other catalogs,
produced a final sample of 278 objects listed in Table 1. Six objects are
shown in Figure 1, displaying the range in surface brightness and absolute
luminosity that the sample entailed.

\section{Observations}

The Second Palomar Sky Survey (PSS-II, see Reid \etal 1991) provided all
the plate material for this project.  The PSS-II differed from the
original sky survey in three important ways.  The first is that the
exposures were made in the latest Kodak IIIa plates, which have greater
resolution and depth than the original surveys 103a emulsions (250 lines
mm$^{-1}$ versus 80 lines mm$^{-1}$).  The smaller grain sizes of the IIIa
emulsions, combined with modern hypersensitization techniques,  also
results in greater sky density and higher uniformity in the plate
background which is critical in the detection of low surface brightness
objects such as dwarf galaxies.

The second difference is that the 48-inch Oschin Schmidt on Palomar
Mountain was significantly modified with a new corrector, an updated mirror
support system and improved plate holders.  The original corrector lens
was replaced in 1985 with a achromatic plate constructed of Schott LLF6
and O'Hara BK7W.  The resulting image sizes were better than 0.5 arcsecs
for in focus point sources.  However, the new corrector has a significant
astigmatic component that required the construction of higher precision
plate holders.  These new plate holders are capable of bending a 1 mm
glass plate of 14 inches in length to an accuracy of $\pm$ 15 $\mu$m and
reduce the astigmatism to less than the typical seeing disk over the
entire field of view.

The\ third difference between the old survey and PSS-II is that the
spacing of the survey fields was decreased from 6 degrees to 5 degrees.
This results in a greater overlay zone on each plate and provides a
greater total plate area to confirm the reality of detected objects.
During this survey, all objects in the overlap fields were independently
confirmed in the neighboring plates.

The plates used for this project are A or B grade, selected for good
surface brightness depth and covering declination zones of the sky that
can be observed with the 305m Arecibo radio telescope.  Plates with a
grade of A are accepted for survey production having excellent uniformity
and depth as well as prefect image quality.  Plates with a grade of B are
rejected for survey production for cosmetic reasons (aircraft or satellite
trails, mild stellar elongation at the plates edges).  B grade plates have
the same depth and uniformity as A grade plates and the differences are
irrelevant for this project.  Due to the limited declination range of the
Arecibo facility, and the incomplete status of the Second Sky Survey, only
35 fields were examined.  All the plate inspection was accomplished by one
of us (JMS) over a period of four months.

The inspection procedure consisted of marking the plate in 1 square degree
regions and using a low power eyepiece to search and mark candidate
objects.  The identity of each marked object was cross-correlated to known
galaxies using the POSS transparencies.  The plate scales remained the
same between the old and new survey and the transparencies contain all the
objects from the NGC, IC, UGC and CGCG catalogs, as well as the Arp and
Markarian lists.  The coordinates of each candidate dwarf were taken
directly from the plate material using a fine ruler and the position of
SAO stars.  Accuracy of the coordinates were typically $\pm$15 arcsecs
based on the centering of the optical imaging, which is sufficient for
detection with the Arecibo beam of 3 arcmin diameter.

We observed the dwarf candidates in our list for the HI line at 21 cm with
the Arecibo 305m telescope during the 1992 and 1993 observing season.  All
observations were made with the 21 cm dual-circular feed positioned to
provide a maximum gain (8 K Jy$^{-1}$) at 1400 MHz.  The 2048 channel
autocorrelator was used and the independent, opposite polarized signals
were each divided into two subcorrelators of 512 channels.  In order to
search a larger velocity space, the secondary local oscillators of each
polarization set of subcorrelators were offset on either side of the
standard local oscillator frequency of 260 MHz by 8.75 MHz, allowing a
total velocity coverage of 8000 km s$^{-1}$, a velocity resolution of 8.6
km s$^{-1}$, and some overlap at the band edges.  The observations were
centered on 4000 km s$^{-1}$, which avoided detection of the strong
Galactic hydrogen signal on the low-velocity end, and extended to 8120 km
s$^{-1}$.  Observations were made in the total power mode with 5 minute
ON- source and OFF-source observations.   In most cases, only one 5-minute
ON-source integration was required for detection.  Wherever possible, the
zenith angle was kept less than 14 degrees to minimize the degradation of
the gain.  Details of the HI observations and analysis will be presented
in an upcoming paper (Eder \etal 1997).

Successful detections were later selected for follow-up CCD imaging on the
Hiltner 2.4m telescope located at Michigan-Dartmouth-M.I.T.  (MDM)
Observatory.  We obtained images using either a Thomson 400 $\times$ 576
pixel CCD (0.25 arcsec pixel$^{-1}$) or a Ford-Loral 2048 $\times$ 2048
pixel CCD binned 3$\times$3 (0.51 arcsec pixel$^{- 1}$), with minimal exposure
times of 25 minutes in Johnson $V$ and 15 minutes in Johnson $I$.  Details
of the optical observations and analysis will be presented in an upcoming
paper (Pildis \etal 1997).

\section{Discussion}

From our initial sample of 310 objects, 277 with high quality indices were
observed on the Arecibo 300m telescope at 21-cm out to a velocity of
10,000 km sec$^{-1}$.  Of the galaxies observed, 145 were detected.  The
objects not detected were probably background LSB spirals (with spiral
patterns that were not visible on the plates) or gas-poor dwarfs (dE's).
The loss of gas-poor dwarfs from the sample is unfortunate since the
distribution of dE's outside of a cluster environment is not known or even
if a true dE exists independent of a high local density or nearby
companion (Binggeli, Tarenghi and Sandage 1990).  However, it is not
critical to our attempt to locate low mass test particles at
cosmologically interesting distances.  The catalog of objects are listed
in Table 1 with the object name in column 1, object coordinates (1950) in
column 2, heliocentric velocity from HI detections in column 3,
morphological type in column 4 and any notes or comments in column 5.
Images of all HI detected objects can be found at
http://zebu.uoregon.edu/$\sim$js/dwarf.html.

One of the first issues concerning the catalog was whether it is an
accurate locator of dwarf galaxies.  The original concept of dwarf galaxy
was derived from the identification of small, faint companions to normal
galaxies such as M32 and the dwarfs around M82.  Later schemes emphasized
their unique morphology (irregular), star formation histories and
their impact on galaxy formation theory (see Binggeli 1993 for a review).
Since most dwarf galaxies were faint in luminosity and surface brightness,
it was not too surprising to find these observables reflected in
similar physical attributes, i.e. dwarf galaxies are low in mass and mass
surface density.  Low mass and density imply an origin in low amplitude
fluctuations in the early Universe and, thus, this study's interest in
them as test particles.  This does not diminish intrinsic interest in
their properties since, after a short inspection of any dwarf catalog, it
is obvious that their smaller mass and densities has produced dramatically
different star formation histories as compared to giant or normal
galaxies.

In order to understand the results of a morphologically selected
sample, and test whether a true sample of dwarf galaxies has been selected
by this catalog, we compare the properties of our catalog objects to dwarf
samples in the literature (Schneider \etal 1990) and the properties of the
UGC catalog (Nilson 1973, Huchtmeier and Richter 1989).  The optical and
HI data for this comparison is presented graphically here, but will be
published in our companion papers.  The reader should note that a HI
detection is required for any absolute magnitude or scale length
comparison since the only source of redshift information is at 21-cm.
Optical redshifts would be very telescope time intensive due to the LSB
nature of the sample and that time was better used in photometry portion
of the project.  Therefore, the following analysis is based on a
HI-selected dwarf sample and by necessity avoids gas-poor dwarfs (e.g.
dE's).

A first order definition of the difference between a dwarf galaxy and a
normal or giant galaxy would focus on a comparison of masses.
Historically, mass is not a directly determined quantity in galaxies and,
thus, luminosity is used since the range of mass-to-light of giant
galaxies for a given Hubble type is quite small due to simple past
histories of star formation.  However, a division between dwarf and giant
galaxies by luminosity distorts the vast range of current star formation
rates in dwarf galaxies.  For example, it is also possible for dwarf
galaxies to have relatively high luminosities (e.g. A1212+06, a BCD in the
Virgo cluster).

The luminosities of our sample galaxies and other optical properties are
shown in Figure 2.  The optical data will be presented and further
discussed in Pildis \etal 1997, however, several trends are worth noting
here.  The total luminosities for the sample are, as expected, low with
most of the galaxies having $M_V > -18$ mags.  The central surface
brightnesses are also low with a typical value between 21 and 23 $I$ mags
arcsec$^{-2}$ (approximately 22.5 to 24.5 $B$ mags arcsec$^{-2}$).  For
comparison, the value of a Freeman disk is 21.65 $B$ mags arcsec$^{-2}$.
The scale length from exponential fits are typically less than 2 kpc (a
normal disk galaxy has scale length around 3 kpc, de Jong 1996).  Lastly,
the mean colors of the sample are quite blue ($V-I$ between 0.5 and 1.0),
but this is typical for LSB galaxies (Schombert \etal 1992, McGaugh 1992)
regardless of dwarf or non-dwarf classification.

As noted above, the luminosity of a galaxy is sharply dependent on the
recent star formation history.  On the other hand, separation by scale
size rather than luminosity has been shown to more closely follow
underlying physical properties of a galaxy (De Jong 1996).  With respect
to isophotal metric size, this sample of morphological selected objects is
also well below the mean of any other galaxy catalog.  Figure 3 displays
the distribution of isophotal sizes for our sample of dwarfs and the
general distribution from galaxies in the UGC catalog with HI detections
(UGC HI, Huchtmeier and Richter 1989, i.e. a comparison of HI selected samples).
The isophotal size, $R_{25}$, is the metric radius at the 25 $I$ mag
arcsecs$^{-2}$ isophote measured directly from surface brightness profiles
of the dwarf galaxies.  This is equivalent to the UGC diameters which are
determined from isophotal levels that are approximately 26.5 $B$ mag
arcsecs$^{-2}$, the so-called Holmberg radius (most LSB objects have
$B-I$=1.5 colors).  Figure 3 also demonstrates that this sample is
composed of objects with radii in the lower 25\% of the distribution of
UGC radii.  The mean $R_{25}$ value for UGC HI selected objects is 15 kpc,
versus a mean radius of 4 kpc for our dwarf sample.

There are two additional methods for dividing a LSB sample into dwarfs and
giants: HI mass and HI profile velocity width.  HI mass is a direct
measure of the gas mass in a galaxy, assuming that the molecular material
is a constant fraction of the total gas mass.  For disk galaxies, there is
a fairly good correlation between gas mass and total mass of a galaxy (see
Giovanelli and Haynes 1988).  However, this relationship does not exist
for dwarf galaxies, probably due to the nature of star formation in dwarf
galaxies which can quickly consume large fractions of the gas reserves.
In addition, due to their weak gravitational potential, star formation and
gas heating is much more effective at removing gas from dwarf galaxies as
compared to their giant galaxy counterparts. Thus, the ratio of gas to
total mass is more dependent on a dwarf galaxy's past star formation
history than its initial physical characteristics.

The distribution of HI masses is shown in Figure 4.  Also shown in Figure
4 is the distribution of HI masses from the Schneider \etal (1990) study
of UGC dwarfs and the general distribution of HI masses for all UGC
galaxies regardless of morphological type (Huchtmeier and Richter 1989,
all corrected to $H_o=85$).  Our sample of morphologically selected
objects matches closely the same range of HI masses as the UGC dwarfs.
Both dwarf samples lie on the lower mass range of the UGC HI sample with
masses less than log $M_{HI}$ = 9.5.  The mean UGC HI mass is log $M_{HI}$
= 9.7.  Interestingly, both the Schneider \etal (1990) and PSS-II samples
have identical cutoffs at log $M_{HI}$ = 9.2 to the point where an
operational definition of a dwarf galaxy is one with an HI mass less than
this value (bright elliptical galaxies being a clear violation of this
criteria).  Given the expected decoupling of HI mass to total mass (see
discussion above), the results presented in Figure 4 is surprising.  We
believe that the strongest statement Figure 4 can demonstrate is to show
that a morphologically selected sample of dwarfs (this study and the UGC
dwarf sample) will yield similar distributions of HI masses.

The HI line profile width is a good tool for determining the mass of disk
galaxies, but is less effective with dwarf galaxies since they are not
completely rotationally supported and probably have a significant
component of their velocity width due to bulk stellar motions.  For an
excellent discussion of converting HI profiles to total mass estimates see
Staveley-Smith, Davies and Kinman (1992).  For our needs it is sufficient
to plot the distribution of HI profile widths as a check on the lack of
high width objects in the sample.  This plot is shown in Figure 5 with
comparison data from the Schneider \etal (1990) and UGC HI sample.  As in
Figure 4, the dwarf samples are composed of the lowest width objects as
compared to the total HI sample of galaxies in the UGC.  The PSS-II dwarf
sample has a slightly lower distribution of profile widths compared to the
Schneider \etal (1990) dwarf sample, but this may be due to a selection
effect to detect face-on LSB objects in a visual survey which, in turn,
emphasizes the non-rotational component to the velocity profile.  Only six
of the detected PSS-II dwarfs displayed a double-horned HI profile
indicative of a disk system.  All six also displayed disk and bulge
morphology and spiral features with inspection by deep CCD images.  These
six objects were latter classified as a new type of galaxy, dwarf spirals,
and described in a previous paper (Schombert \etal 1995).

\section{Results}

A sample of isolated dwarf galaxies has been selected from PSS-II plates.
The candidates are selected by morphological criteria and then confirmed
with HI detection.  We show that a sample selected in this fashion yields
an excellent set of isolated dwarf galaxies, (i.e. not companions to
bright galaxies), and our plots and discussion in the previous section
demonstrate that morphologically selected candidates yield a sample
dwarfs, as defined by optical or HI properties.

The redshift distribution of the catalog lies between 500 and 10,000 km
sec$^{-1}$, making it a more useful map of large scale structure than
previous dwarf catalogs. Our primary result is shown in Figure 6, the
redshift cone plot of our dwarf sample as compared to the distribution of
galaxies from the CfA Redshift Survey (ZCAT, see Marzke, Huchra and Geller
1994) and the UGC HI sample.  The ZCAT and UGC samples are selected from
the declination zones +5 to +25$^{\circ}$.

The ZCAT sample was selected for bright galaxies (likely to be higher mass
objects, a test of biased gaalxy formation).  The UGC HI sample was
selected to compare our HI dwarf sample's redshift distribution with
another HI selected sample.  Neither the ZCAT nor the UGC HI distribution
is visually any different from the redshift distribution of our dwarf
galaxies, in conflict with the predictions of biased galaxy formation
theory.  A full analysis of the data will be presented in the latter
papers of our series.

\acknowledgements

We wish to thank the generous support of the University of Michigan and
the Michigan- Dartmouth-M.I.T Observatory in carrying out the photometry
portion of this program and Arecibo Observatory for the allocation of time
to search for HI emission from the candidate dwarf galaxies.  The research
described herein was carried out by the Jet Propulsion Laboratory,
California Institute of Technology, under a contract with the National
Aeronautics and Space Administration.  This work is based on photographic
plates obtained at the Palomar Observatory 48-inch Oschin Telescope for
the Second Palomar Observatory Sky Survey which was funded by the Eastman
Kodak Company, the National Geographic Society, the Samuel Oschin
Foundation, the Alfred Sloan Foundation, the National Science Foundation
grants AST 84-08225 and AST 87-18465, and the National Aeronautics and
Space Administration grants NGL 05002140 and NAGW 1710.  R.A.P. was
partially supported by a National Science Foundation Graduate Fellowship.
This research has made use of the NASA/IPAC Extragalactic Database (NED)
which is operated by the Jet Propulsion Laboratory, California Institute
of Technology, under contract with the National Aeronautics and Space
Administration.

\clearpage

\figcaption{CCD images for six dwarfs from our catalog.  North is at the
top, East to the left.  Each frame is 8 kpc on a side.  Images of all the
HI detected dwarfs are available on the Dwarf Galaxy Survey homepage
(http://zebu.uoregon.edu/$\sim$js/dwarf.html).}

\figcaption{The optical properties of the HI detected dwarf galaxies are
shown (data from Pildis \etal 1997).  The optical properties are similar
to previous dwarf samples being low in luminosity, small in scale length
and faint in central surface brightness.  The bluer $B-I$ colors (a
typical elliptical has a $B-I$=1.5) is common for LSB galaxies (Schombert
\etal 1992).}

\figcaption{The distribution of isophotal radii (radius at 25.0 $I$ mag
arcsec$^{-2}$) are shown for our dwarf sample and the total HI UGC catalog
for all Hubble types.  The dwarf galaxies in this catalog are all less
than 25\% of the typical size for UGC objects.}

\figcaption{The distribution of HI masses is shown for our dwarf sample,
the UGC dwarf sample (Schneider \etal 1990) and the complete UGC sample.
Our sample of morphologically selected objects matches closely the same
range of HI masses as the UGC dwarfs.  Both dwarf samples lie on the lower
mass range of the UGC HI sample with masses less than log $M_{HI}$ = 9.5.
The mean UGC HI mass is log $M_{HI}$ = 9.7.} 

\figcaption{The distribution of HI profile widths is shown for our dwarf
sample, the UGC dwarf sample (Schneider \etal 1990) and the complete UGC
sample.   As in Figure 4, the dwarf samples are composed of the lowest
width objects as compared the total HI sample of galaxies in the UGC.}

\figcaption{The redshift cone plot of our dwarf sample as compared to the
distribution of galaxies from the CfA Redshift Survey (ZCAT, see Marzke,
Huchra and Geller 1994) and the UGC HI sample.  The ZCAT and UGC samples
are selected from the declination zones +5 to +25$^{\circ}$.  The ZCAT
sample was selected for bright galaxies (likely to be higher mass
objects).  The UGC HI sample was selected to test for selection effects
due to an HI selected sample.  Neither distribution is visually any
different from the redshift distribution of our dwarf galaxies, in
conflict with the predictions of biased galaxy formation theory.}


\pagestyle{empty} \clearpage
\begin{figure} 
\plotfiddle{fig1.ps}{11.5truein}{0}{100}{100}{-310}{170} \end{figure} 
\clearpage \begin{figure}
\plotfiddle{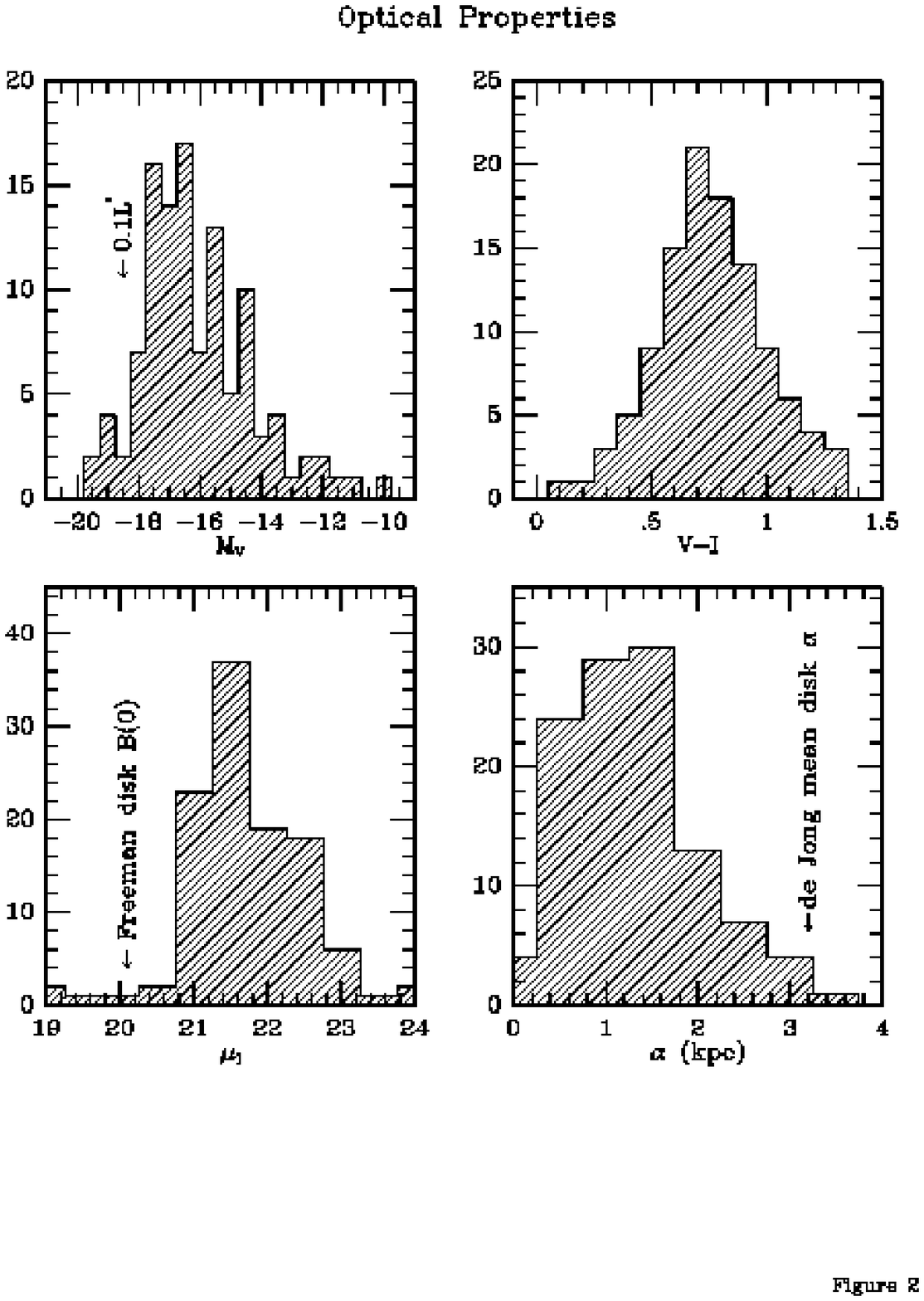}{11.5truein}{0}{100}{100}{-310}{170} \end{figure}
\clearpage \begin{figure}
\plotfiddle{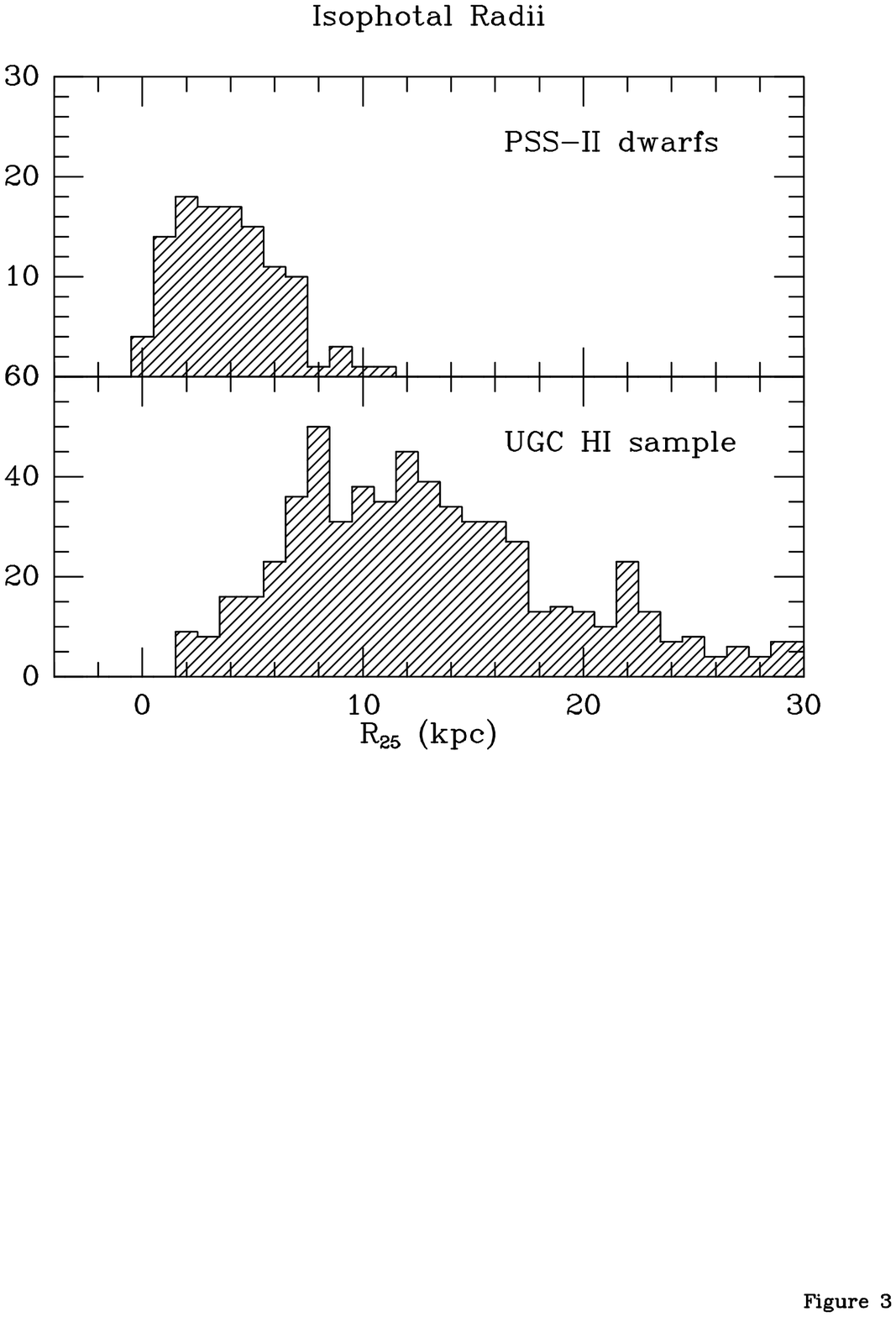}{11.5truein}{0}{100}{100}{-310}{170} \end{figure}
\clearpage \begin{figure}
\plotfiddle{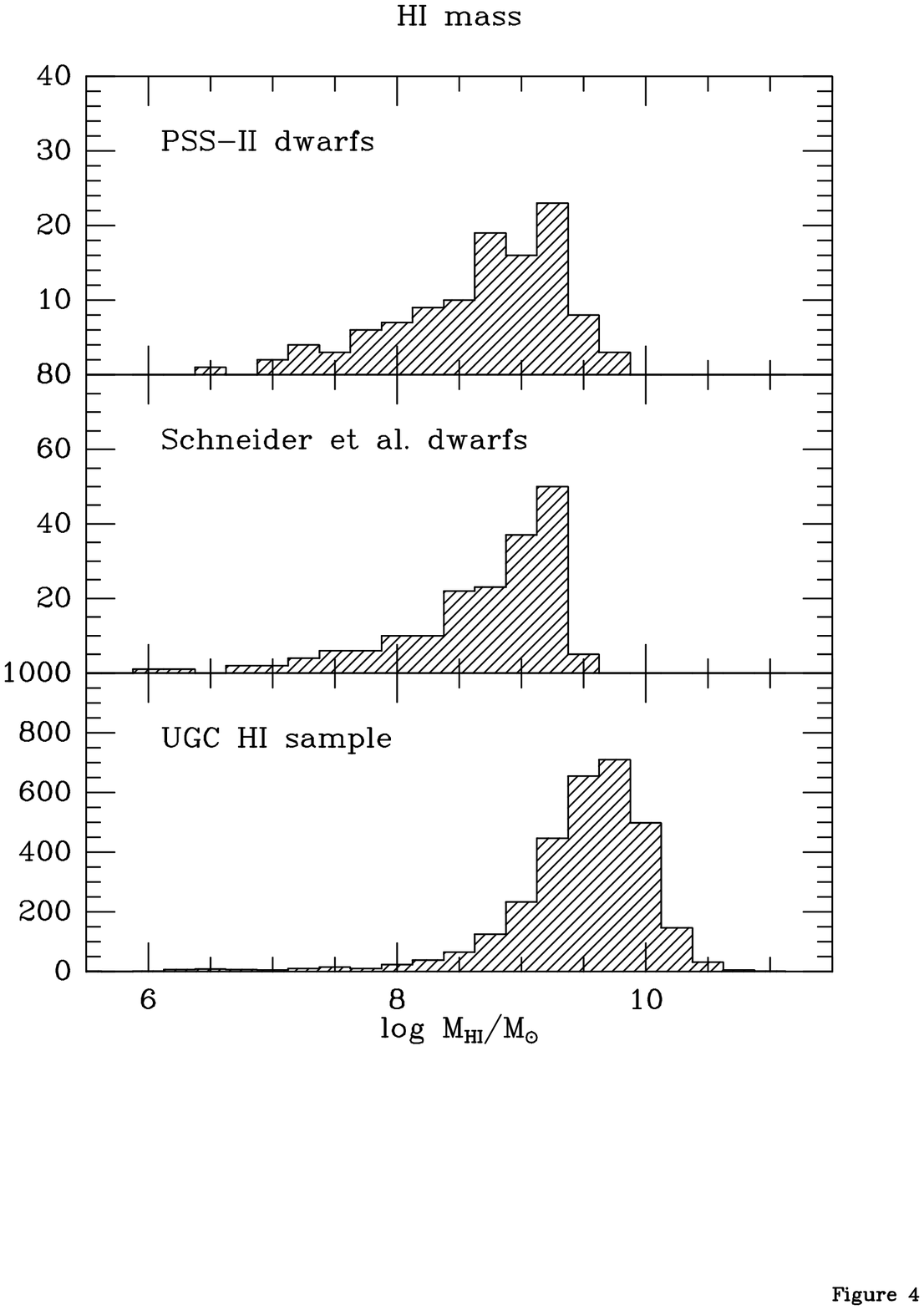}{11.5truein}{0}{100}{100}{-310}{170} \end{figure}
\clearpage \begin{figure}
\plotfiddle{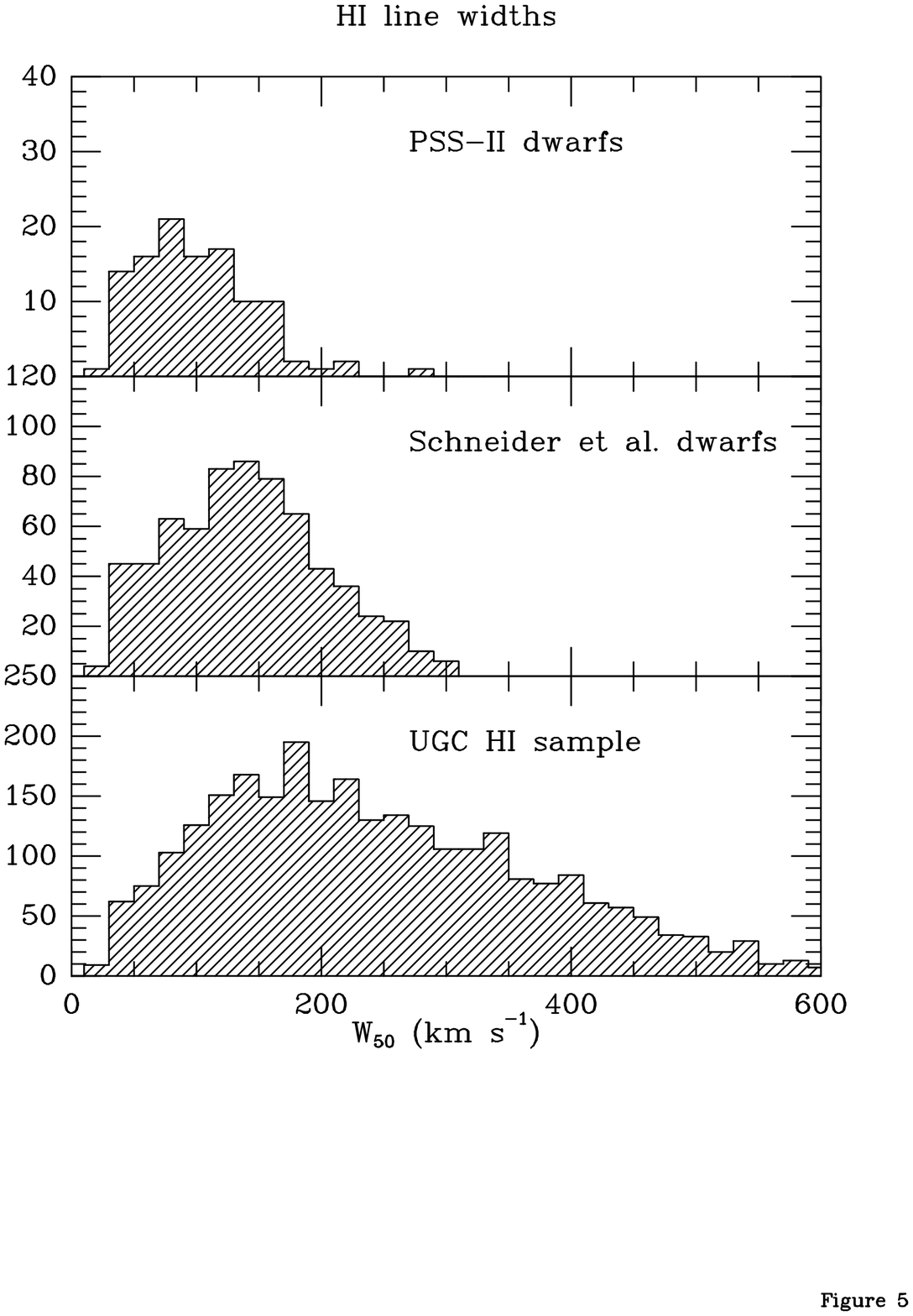}{11.5truein}{0}{100}{100}{-310}{170} \end{figure}
\clearpage \begin{figure}
\plotfiddle{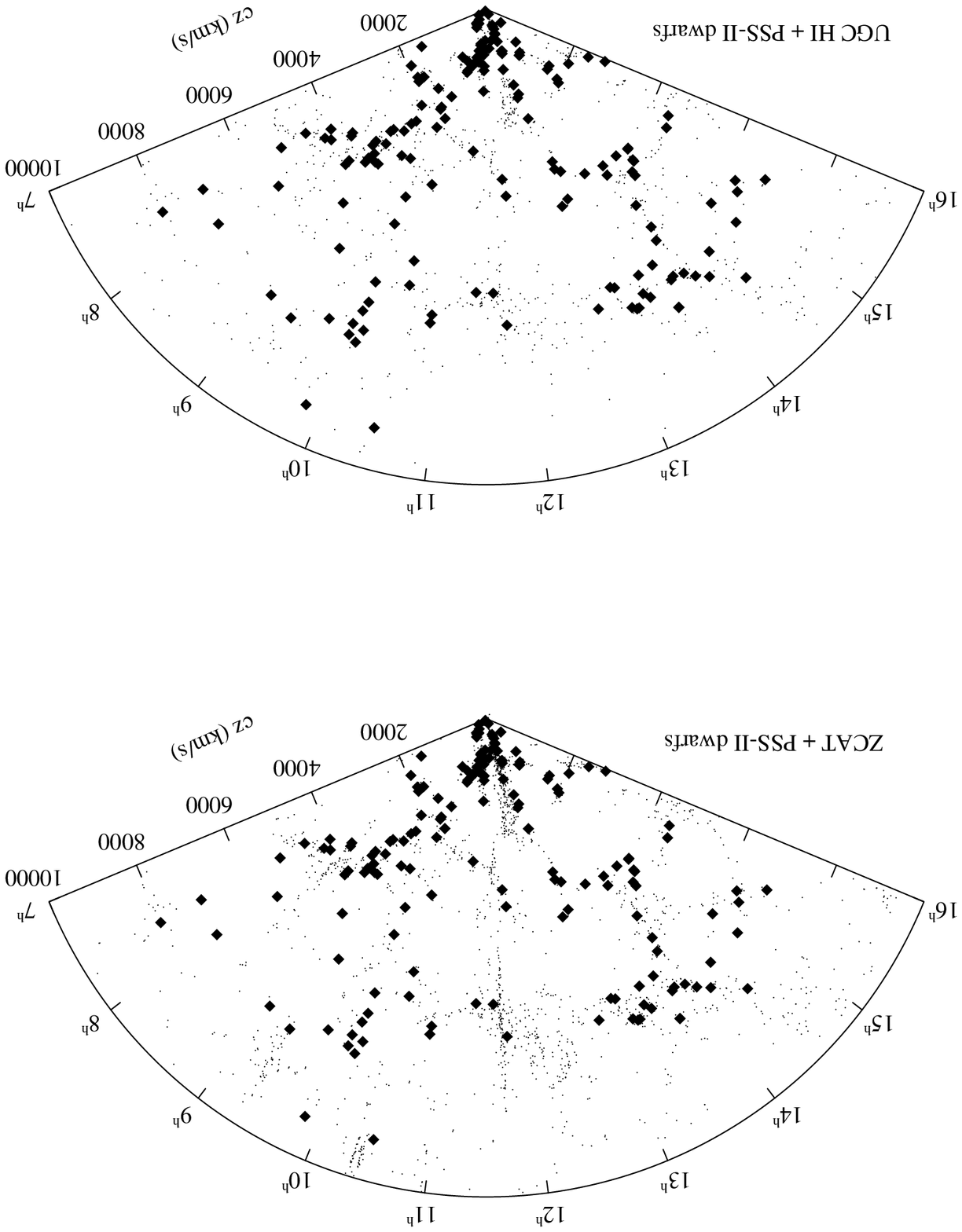}{11.5truein}{0}{100}{100}{-310}{130} \end{figure}

\end{document}